\documentclass[letterpaper,english,reprint, aps]{revtex4-1}
\usepackage[T1]{fontenc}
\usepackage[latin9]{inputenc}
\usepackage{textcomp}
\usepackage{mathrsfs}
\usepackage{amsmath}
\usepackage{amssymb}
\usepackage{graphicx}

\makeatletter

\newcommand*\LyXThinSpace{\,\hspace{0pt}}
\pdfpageheight\paperheight
\pdfpagewidth\paperwidth

\makeatother

\usepackage{babel}
\begin{document}
\title{Thermodynamic coupling rule for far-from-equilibrium systems}
\author{Shanhe Su}
\address{Department of Physics, Xiamen University, Xiamen 361005, People's
Republic of China.}
\author{Wei Shen}
\address{Department of Physics, Xiamen University, Xiamen 361005, People's
Republic of China.}
\author{Jianying Du}
\address{Department of Physics, Xiamen University, Xiamen 361005, People's
Republic of China.}
\author{Jincan Chen}
\email{jcchen@xmu.edu.cn}

\address{Department of Physics, Xiamen University, Xiamen 361005, People's
Republic of China.}
\date{\today}
\begin{abstract}
The energy conversion efficiency of far-from-equilibrium systems is
generally limited by irreversible thermodynamic fluxes that make contact
with different heat baths. For complex systems, the states of the
maximum efficiency and the minimum entropy production are usually
not equivalent. Here we show that the proper adjustments of the interaction
between the energy and matter currents offer some important criteria
for the performance characterizations of thermal agents, regardless
of the system types and transition protocols. The universal thermodynamic
coupling rule plays a critical role in irreversible processes. A double
quantum dot system is applied to demonstrate that the performances
of heat engines or refrigrators can be enhanced by suitably adjusting
the coupling strength between thermodynamic fluxes. 

\end{abstract}
\maketitle
For open inhomogeneous systems, the thermodynamic process driven by
a temperature difference between two heat baths is generally irreversible.
At a stationary state, the entropy production rate $\sigma$ is greater
than or equal to zero, which is exactly balanced by the entropy flow
through the terminals \citep{key-1,key-2,key-3}. If $\sigma=0$,
the performance of the engine (refrigerator) reaches its limit and
the efficiency {[}coefficient of performance (COP){]} will be equal
to the Carnot value \citep{key-4,key-5,key-6}. However, since all
complex natural processes are irreversible, the minimum entropy production
rates have been found inefficiently to ensure the maximum efficiency
and COP \citep{key-7,key-8}. 

Pietzonka et al. proved that a universal trade-off among the power,
efficiency, and constancy exists in non-equilibrium heat engines \citep{key-9,key-10}.
Whitney found that the maximum efficiency of an irreversible device
occurs when the system lets all particles go through in a certain
transmission window, but none at other energies \citep{key-11}. Benenti
et al. pointed out that the upper bound of the maximum efficiency
of a thermoelectric generator is reached when the figure of merit
$ZT\rightarrow\infty$. This limit corresponds to the tight coupling
condition, for which the Onsager matrix becomes singular and therefore
the ratio of the heat current to the electric current is independent
of the applied thermodynamic forces \citep{key-12,key-13,key-14}.
The strong coupling condition can be obtained for a nanothermoelectric
device consisting of a single quantum level \citep{key-15,key-16}.
More generally, the dissipative thermodynamic fluxes rely on the temperature
and chemical potential gradients \citep{key-17,key-18,key-19,key-20}.
It is unlikely to identify two quantities involving proportional relationships.
The question arises as to whether there are universal principles relevant
to the thermodynamic fluxes and forces that provide ways to the maximum
efficiency and COP of far-from-equilibrium systems.

Here, we present a novel criterion for attaining the upper bound for
the efficiency (COP) of any heat engine (refrigerator) running between
two heat baths. Our main result is that the connection between the
energy and matter currents is suitable for the performance characterization
of thermal agents regardless of the system types and transition protocols.
Tracing from the master equation, we start with the analysis of the
entropy and the statistics of energy and matter transfers. Results
will show that the most convenient optimization technique is to alter
the coupling strength of thermodynamic fluxes. An interacting double
quantum dot (DQD) system will be used to verify our predictions.

For a non-degenerate system weakly coupled to different environmental
modes, the dynamics of the populations satisfies a rate equation 
\begin{equation}
\dot{P}{}_{i}=\sum_{j}\mathscr{L}_{ij}P_{j}=\sum_{j}\sum_{v}\mathscr{L}_{ij}^{(\nu)}P_{j},
\end{equation}
where $P_{i}$ is the probability of finding the system in the eigenstate
$\left|i\right\rangle $, $\mathscr{L}_{ij}$ describes the transfer
rate from state $j$ to state $i$, and $\mathscr{L}_{ij}^{(\nu)}$
is the rate contributed by the coupling with the reservoir $\nu$.
All possible decays of $P_{i}$ are contained within the factor $\sum_{i}\sum_{v}\mathscr{L}_{ii}^{(\nu)}$.
When an open inhomogeneous system obeying Eq. (1) is placed in contact
with the hot and cold reservoirs (denoted H and C), the entropy production
rate due to irreversible processes reads {[}Supplementary Eq. (S6){]}

\begin{align}
\sigma & =-\beta_{H}J^{(H)}-\beta_{C}J^{(C)}\nonumber \\
 & =\left(\beta_{C}-\beta_{H}\right)I_{E}^{\left(H\right)}+\left(\beta_{H}\mu_{H}-\beta_{C}\mu_{C}\right)I_{M}^{\left(H\right)},\label{eq:2p}
\end{align}
where $J^{\left(\nu\right)}=I_{E}^{\left(\nu\right)}-\mu_{\nu}I_{M}^{\left(\nu\right)}$
describes the heat current entering the system from reservoir $\nu$,
$\beta_{\nu}$ represents the inverse value of the temperature, and
$\mu_{\nu}$ is the chemical potential. The energy current $I_{E}^{\left(\nu\right)}$
and the matter current $I_{M}^{\left(\nu\right)}$ associated with
reservoir $\nu$ are positive variables when flowing into the the
system. The second formula in Eq. (2) has been expressed in terms
of the currents entering the system from the hot reservoir, because
both energy and matter currents are conserved within the system, i.e.,
$I_{E}^{\left(H\right)}+I_{E}^{\left(C\right)}=0$ and $I_{M}^{\left(H\right)}+I_{M}^{\left(C\right)}=0$.
On the basis of Eq. (\ref{eq:2p}), we are able to determine the universal
thermodynamic coupling rule for the systems interacting with two terminals.
Consider two reservoirs with the same temperature but at different
electrochemical potentials $\mu_{C}-\mu_{H}=eV\geq0$, where $e$
is the positive elementary charge. The transfer of an electron from
C to H increases the entropy of the universe by $\beta eV$ independent
of the electron energy. For the general case where $\beta_{H}<\beta_{C}$
and $\mu_{H}<\mu_{C}$, an irreversible process yields $\left(\beta_{C}-\beta_{H}\right)I_{E}^{\left(H\right)}+\left(\beta_{H}\mu_{H}-\beta_{C}\mu_{C}\right)I_{M}^{\left(H\right)}\geq0$
(Section 1 in Supplementary). When electrons spontaneously flow from
H to C, the system behaves as a heat engine. Under this condition,
$I_{M}^{\left(H\right)}>0$ and the ratio of the energy and matter
currents is bounded by
\begin{equation}
I_{E}^{\left(H\right)}/I_{M}^{\left(H\right)}\geq\varTheta,\label{eq:3p}
\end{equation}
where $\varTheta=\left(\beta_{C}\mu_{C}-\beta_{H}\mu_{H}\right)/\left(\beta_{C}-\beta_{H}\right)$.
The efficiency of the thermal engine expresses the fraction of heat
that becomes useful work, i.e.,

\begin{equation}
\eta=\frac{J^{H}+J^{C}}{J^{H}}=\frac{\mu_{C}-\mu_{H}}{I_{E}^{\left(H\right)}/I_{M}^{\left(H\right)}-\mu_{H}}.\label{eq:4p}
\end{equation}
If $I_{M}^{\left(H\right)}<0$, electrons flow spontaneously from
C to H, implying that the ratio $I_{E}^{\left(H\right)}/I_{M}^{\left(H\right)}$
has an upper bound given by
\begin{equation}
I_{E}^{\left(H\right)}/I_{M}^{\left(H\right)}\leq\varTheta.\label{eq:5p}
\end{equation}
At this juncture, the system is capable of removing heat from the
lower-temperature reservoir. The COP of the refrigerator is the ratio
of the heat removed from C to the input work, i.e.,
\begin{equation}
\phi=\frac{J^{C}}{J^{H}+J^{C}}=\frac{I_{E}^{\left(H\right)}/I_{M}^{\left(H\right)}-\mu_{C}}{\mu_{C}-\mu_{H}}.\label{eq:6p}
\end{equation}
Equations (3-6) reveal that the coupling strength $I_{E}^{\left(H\right)}/I_{M}^{\left(H\right)}$
plays an important role in determining the efficiency and COP of the
thermal devices regardless of the system complexity. The interplay
between the energy and matter currents characterizes the energy conversion
performance of physical and biological systems. $\eta$ $\left(\phi\right)$
is a linear decreasing (increasing) function of $I_{E}^{\left(H\right)}/I_{M}^{\left(H\right)}$
at given values of $\mu_{C}$ and $\mu_{H}$. The criterion function
$\varTheta$ simply relies on a set of reservoir parameters. This
is the main result of the present work: by suitably adjusting the
coupling strength between the energy and matter currents, electrons
can be transferred nearly reversibly between two reservoirs with arbitrary
temperatures and electrochemical potentials. 

For a quantum dot with a single transition frequency embedded between
two fermionic junctions, the energy and matter currents exhibit tight
coupling properties \citep{key-15}. Every electronic jump carries
the same amount of energy $\varepsilon$, yielding the perfect couple
situation $I_{E}^{\left(H\right)}=\varepsilon I_{M}^{\left(H\right)}$.
At steady state, the entropy production rate turns into $\sigma=\left[\left(\beta_{C}-\beta_{H}\right)\varepsilon+\left(\beta_{H}\mu_{H}-\beta_{C}\mu_{C}\right)\right]I_{M}^{\left(H\right)}$,
which is directly proportional to the matter current. Supposing $\varepsilon=\varTheta$,
we are able to find $\sigma=0$ and obtain a reversibe nanothermoelectric
device with the Carnot efficiency or COP. At this particular transition
energy, the temperature and electrochemical driving forces cancel
each other out, since the two reservoirs behave as if they were in
thermodynamic equilibrium \citep{key-5}. The probability densities
for electrons are equal on both sides of the transport channel. This
is, however, not the case for a multilevel system, as the transferred
energy depends on a specific type of jumping. In general, the energy
and matter currents are not in direct proportion and cannot simultaneously
tend to zero, resulting in a positive entropy production rate. Maximum
efficiency may not correspond to a minimal entropy production for
irreversible processes.

\begin{figure}
\includegraphics[scale=0.35]{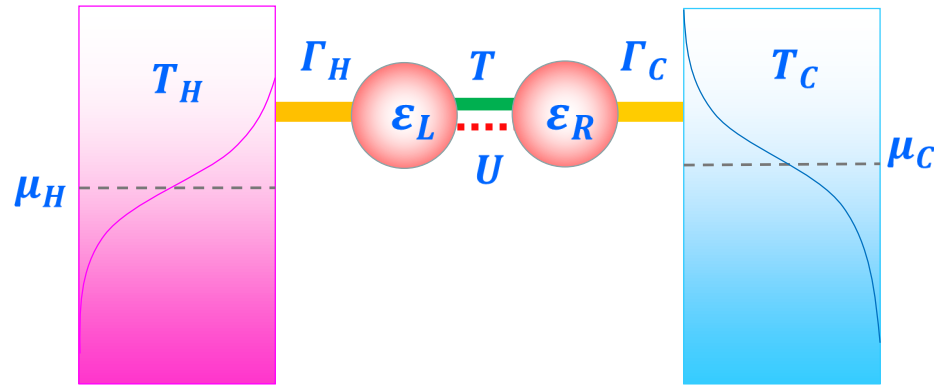}\caption{The schematic representation of a double quantum dot (DQD) system
weakly coupled to two fermionic reservoirs via the rates $\Gamma_{H/C}$
(origin solid lines). The energy of the left dot (resp. the right
dot) with one-electron occupation is equal to $\varepsilon_{L}$ (resp.
$\varepsilon_{R}$). The electron hops between the left and right
dots with internal tunneling amplitude $T$ (green solid line) and
Coulomb interaction strength $U$ (red dotted line). }
\end{figure}
To reveal the potential application of the thermodynamic coupling
rule mentioned above, we build a DQD sytem that is weakly coupled
to two separate electronic baths. The DQD (Fig. 1) is modeled by the
Hamiltonian
\begin{equation}
H_{S}=\varepsilon_{L}d_{L}^{\dagger}d_{L}+\varepsilon_{R}d_{R}^{\dagger}d_{R}+T\text{\ensuremath{\left(d_{L}d_{R}^{\dagger}+d_{R}d_{L}^{\dagger}\right)}}+Ud_{L}^{\dagger}d_{L}d_{R}^{\dagger}d_{R},\label{eq:p7}
\end{equation}
where $T$ denotes the inter-dot tunneling, and $U$ is the Coulomb
interaction \citep{key-21,key-22}. $d_{L\left(R\right)}^{\dagger}$
creates one electron on the $L\left(R\right)$ QD with energy $\varepsilon_{L(R)}$.
The Hamiltonian of the fermionic baths (H and C) is given by

\begin{equation}
H_{B}=\mathop{\sum}_{k}\varepsilon_{kH}c_{kH}^{\dagger}c_{kH}+\mathop{\sum}_{k}\varepsilon_{kC}c{}_{kC}^{\dagger}c_{kC}\text{.}
\end{equation}

The interaction between the DQD and the environment reads

\begin{align}
H_{I} & =\mathop{\sum}_{k}\left(t_{kH}d_{L}c_{kH}^{\dagger}+t_{kL}^{\ast}c_{kH}d_{L}^{\dagger}\right)\nonumber \\
 & +\mathop{\sum}_{k}\left(t_{kC}d_{R}c_{kC}^{\text{\dag}}+t_{kC}^{\text{\textasteriskcentered}}c_{kC}d_{R}^{\text{\dag}}\right),\label{eq:15}
\end{align}
where $t_{kH\left(C\right)}$ denote the coupling strengths of the
transitions between the DQD and the hot (cold) electronic reservoir.
The eigenstates of $H_{S}$ are given by the tensorial product of
the individual QD states, i.e., $\left|0\right\rangle $ and $\left|1\right\rangle $.
In the increasing energy order, we have the four eigenstates labeled
as $\left|v_{0}\right\rangle =\left|00\right\rangle $, $\left|v_{-}\right\rangle \propto\left[-\left(\Delta+\sqrt{\Delta^{2}+T^{2}}\right)\left|10\right\rangle +T\left|01\right\rangle \right]$,
$\left|v_{+}\right\rangle \propto\left[-\left(\Delta-\sqrt{\Delta^{2}+T^{2}}\right)\left|10\right\rangle +T\left|01\right\rangle \right]$,
and $\left|v_{2}\right\rangle =\left|11\right\rangle $, where $\Delta=(\varepsilon_{R}-\varepsilon_{L})/2$
and $\varepsilon=(\varepsilon_{R}+\varepsilon_{L})/2$. The respective
eigenvalues are $E_{0}=0$, $E_{-}=\varepsilon-\sqrt{\Delta^{2}+T^{2}}$,
$E_{+}=\varepsilon+\sqrt{\Delta^{2}+T^{2}}$, and $E_{2}=2\varepsilon+U$.
Using the fundamental density matrix equation describing dynamics
of the DQD, we readily obtain the energy current $I_{E}^{\left(\nu\right)}$
and the electron current $I_{M}^{\left(\nu\right)}$ flowing into
the the system from the two leads (as shown in Supplementary in detail).
From Eqs. (\ref{eq:4p}) and (\ref{eq:6p}), and Eqs. (S9) and (S10)
in the Supplementary, the thermodynamic irreversibility and the optimal
performances of the DQD system will be estimated. 

\begin{figure}
\includegraphics[scale=0.38]{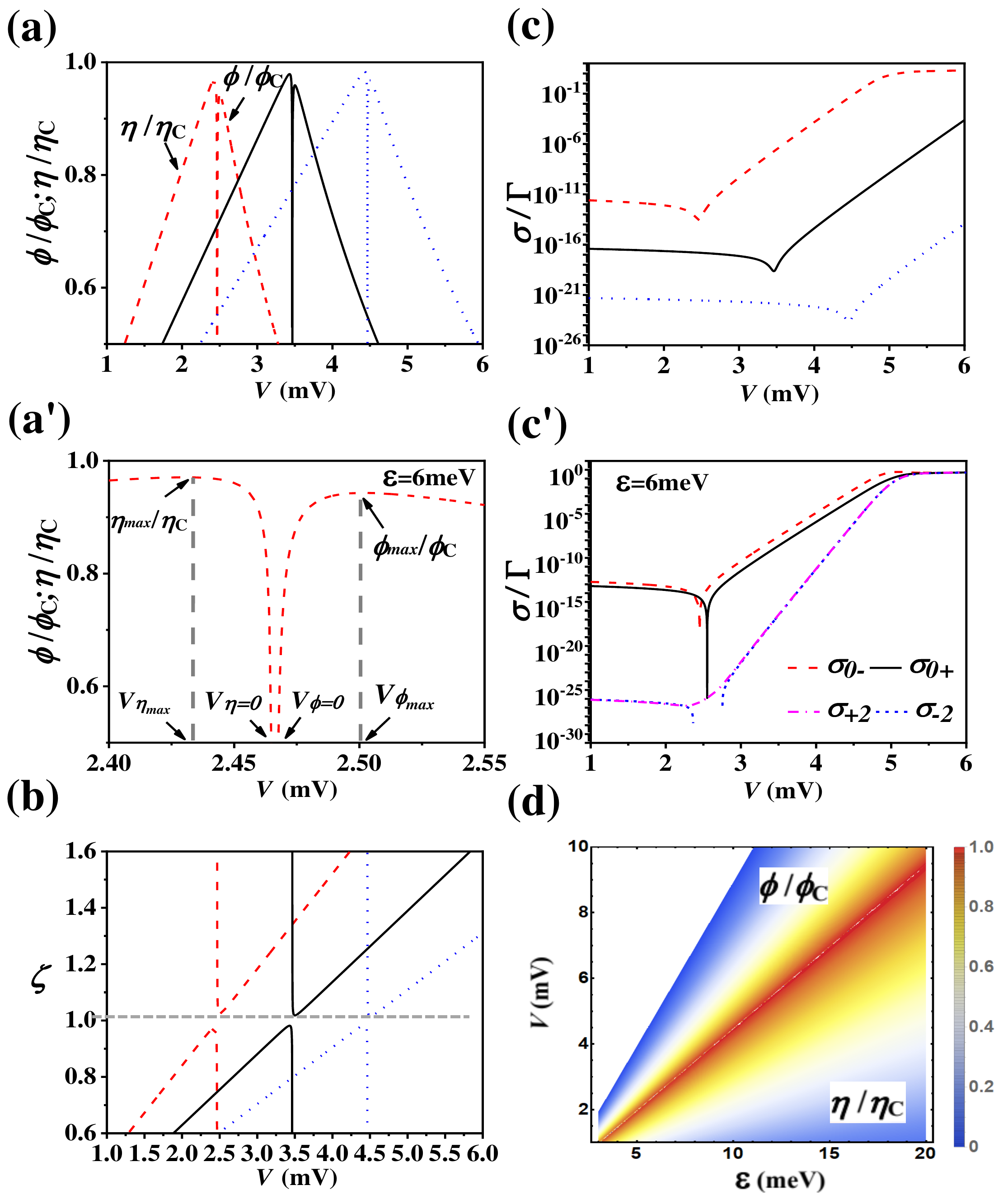}\caption{(a) The efficiency $\eta/\eta^{C}$ and COP $\phi/\phi^{C}$ (normalized
to the Carnot value) of the non-equilibrium DQD system, (b) the dimensionless
coupling parameter $\zeta=\varTheta I_{M}^{\left(H\right)}/I_{E}^{\left(H\right)}$,
and (c) the entropy production rate $\sigma$ as functions of the
voltage $V$ at  $\varepsilon=6meV$ (red dashed line), $8meV$ (black
solid line), $10meV$(blue dotted line). Figure 2(a') shows $\eta/\eta^{C}$
and $\phi/\phi^{C}$ versus $V$ in the range from 2.4 to 2.55mV at
$\varepsilon=6meV$. Figure 2(c') displays the entropy production
rate $\sigma_{ab}$ contributed by the transition from state $b$
to state $a$ at $\varepsilon=6meV$. (d) The contour plot of the
efficiency and COP as functions of $\varepsilon$ and $V$. The parameters
$\beta_{H}=0.5eV$, $\beta_{C}=1eV$, $\Delta=0meV$, $\mu_{H}=1meV$,
$\mu_{C}=\mu_{H}+eV$, $U=T=0.1meV$, and $\Gamma_{H}=\Gamma_{C}=\Gamma$.}
\end{figure}
All thermal devices between two heat reservoirs are less efficient
than a reversible Carnot cycle operating between the same reservoirs.
The corresponding Carnot values of reversible heat engines and refrigerators
are $\eta^{C}=1-\beta_{H}/\beta_{C}$ and $\phi^{C}=\beta_{H}/\left(\beta_{C}-\beta_{H}\right)$,
respectively. Figure 2 (a) shows that the curves of the normalized
efficiency $\eta/\eta^{C}$ and COP $\phi/\phi^{C}$ belong to two
different ridges. In the left ridge, where the chemical driving force
is small, the system lies in the regime suitable for heat moving from
the hot side to the cold side, while the electron flow converts directly
a part of heat into electricity. The system operates as a heat engine
and the normalized efficiency $\eta/\eta^{C}$ is shown on the plot
for this range. According to Eq. (4) and Fig. 2(b), the dimensionless
coupling parameter $\zeta=0$ {[}i.e., $I_{E}^{\left(H\right)}/I_{M}^{\left(H\right)}\rightarrow\infty${]}
at $V=V_{\eta=0}$, which is the value of the voltage at the zero
efficiency {[}Fig. 2(a'){]}. The electrochemical potential gradient
becomes dominant in the process of transferring electrons with the
increase of the voltage and pumps the heat against the thermal gradient.
Therefore, the right ridge plots the normalized COP of a refrigerator
$\phi/\phi^{C}$ versus $V$. There exists a point $V=V_{\phi=0}$
such that $\phi/\phi^{C}=0$ and $I_{E}^{\left(H\right)}/I_{M}^{\left(H\right)}=\mu_{C}$,
as indicated in Eq. (6) and Fig. 2(a'). In the regions of $V\leq V_{\eta=0}$
and $V\geq V_{\phi=0}$, one can find a maximum efficiency $\eta_{max}/\eta^{C}$
whose corresponding voltage is $V_{\eta_{max}}$, and a maximum COP
$\phi_{max}/\phi^{C}$ whose corresponding voltage is $V_{\phi_{max}}$
{[}Fig. 2(a'){]}. In the region between $V=V_{\eta=0}$ and $V_{\phi=0}$,
the system cannot work as a heat engine or a refrigerator. The gap
between refrigerator and heat engine modes in absence of tight coupling
has also been observed beyond the Born\textendash Markov approximation
\citep{key-23,key-24}.

Figures 2(a) and (b) demonstrate that the characteristics of the DQD
system is inseparably related to the coupling between the energy and
matter currents. The normalized efficiency and COP have perfect positive
correlation with the dimensionless coupling parameter $\zeta$. When
$I_{E}^{\left(H\right)}/I_{M}^{\left(H\right)}$ is tight coupling
to the critial value $\varTheta$ given by the reservoir properties,
$\zeta$ approaches unity, and $\eta/\eta^{C}$ (COP) reaches its
extreme value at $V_{\eta_{max}}$ ($V_{\phi_{max}}$). For the DQD
system, the nonlinear relationship makes $I_{E}^{\left(H\right)}$
and $I_{M}^{\left(H\right)}$ unable to be zero at the same time.
This difference allows $\zeta$ to have one-sided limit that equals
negative or positive infinity as $I_{E}^{\left(H\right)}$approaches
0. In the limit of a single transition channel, electron and heat
flows are perfectly coupled, since every single electron carries exactly
the same amount of energy. Under this circumstance, we are capable
of establishing reversible electronic energy transfer between the
reservoirs at the point of $I_{E}^{\left(H\right)}/I_{M}^{\left(H\right)}=\varTheta$. 

The energy-matter coupling condition offers some important criteria
for determining the performance of non-equilibrium systems. In order
to approach the Carnot efficiency, the thermodynamic processes must
be reversible and involve no change in entropy. This means that the
Carnot performance is an idealization, because no real processes are
reversible. Figure 2 (c) illustrates the entropy production rate $\sigma$
as a function of the voltage $V$. Irreversible process is accompanied
by the flows of matter and energy, which always results in some increase
in the entropy. Figure 2(c') points out that the entropy production
rate is mainly dependent upon the transition from state $\left|v_{0}\right\rangle $
to state $\left|v_{+}\right\rangle $ and from state $\left|v_{0}\right\rangle $
to state $\left|v_{-}\right\rangle $, becasue the population of electrons
is an exponentially decreasing function of the transition energy.
In the area between $V_{\eta=0}$ and $V_{\phi=0}$, $\sigma$ is
actually quite small, but the device is not able to create a power
and cooling cycle. The entropy production rate in term of an engine
efficiency can be written as $\sigma=J^{(H)}\beta_{C}\left[\eta^{C}-\eta\right]$.
Obviously, when $\sigma=0$, the efficiency $\eta$ can be shown to
be equal to the Carnot limit $\eta^{C}$. However, $\sigma\neq0$
for irreversible thermodynamic processes in complex systems. The working
conditions of the maximum efficiency and the minimum entropy production
are usually not equivalent, because the relation between $\sigma$
and $\eta$ may not be monotonic in mathematics {[}indicated in Figs.
2 (a) and (c){]}. At the beginning, the efficiency increases with
$V$. The efficiency attains its maximum at $V=V_{\eta_{max}}$. In
the region of $V_{\eta_{=0}}>V>V_{\eta_{max}}$, $\sigma$ and $\eta$
simultaneously decreases with $V$. In addition, it can be observed
from Eqs. (\ref{eq:2p}) and (\ref{eq:4p}) that $\sigma$ and $\eta$
can both diminish by reducing $I_{M}^{\left(H\right)}$, when $I_{E}^{\left(H\right)}$is
constant or changes slowly. The same phenomenon could be observed
in driving an irreversible Carnot refrigerator. Having a minimal amount
of $\sigma$ does not mean having the largest COP. 

The proper adjustment of the interaction between the energy and matter
currents provides an effective way to enhance the performance of non-equilibrium
systems. Figure 2(d) depicts the optimization of the efficiency and
COP with respect to the energy level $\varepsilon$ and the voltage
$V$. For a small $\varepsilon$, the electric field is the main driving
force for the heat transfer. At this juncture, a positive thermal
energy leaves the cold reservoir, and a positive thermal flux enters
the hot reservoir.  As $\varepsilon$ rises, heat will spontaneously
travel from the hot reservoir to the cold reservoir. The DQD system
allows the direct conversion of heat to electricity. A clear boundary
marks an abrupt shift between the working regions of a heat engine
and a refrigerator. This means that the heat engine or refrigerator
cannot work in the region of the abrupt shift, i.e., $V\in\left[V_{\eta_{=0}},V_{\phi_{=0}}\right]$
and $\varepsilon\in\left[\varepsilon_{\phi_{=0}},\varepsilon_{\eta_{=0}}\right]$,
where $\varepsilon_{\phi=0}$ is the value of $\varepsilon$ at the
zero COP and $\varepsilon_{\eta=0}$ is the value of $\varepsilon$
at the zero efficiency. We also observed that the more the coupling
parameter $I_{E}^{\left(H\right)}/I_{M}^{\left(H\right)}$approaches
$\varTheta$ for a given QD energy level, the better the performance
of the heat engine and refrigerator (Section 3 in Supplementary).
In a practical design, we should make our best effort to match the
energy-matter coupling condition. When $\varepsilon>>U$ and $\varepsilon>>T$,
all electrons are tunneling at energy $\varepsilon$. One then recovers
the optimal performance approaching Carnot efficiency or COP {[}as
indicated in Fig. 2 (d){]}, because the system behaves like a thermoelectric
machine consisting of a single quantum level \citep{key-5,key-13,key-15}.
Similar phenomena can also be observed under the condition that the
inter-dot tunneling amplitude and the electrostatic force are incredibly
larger than the QD energy levels, where the two lowest electronic
states $\left|v_{0}\right\rangle $ and $\left|v_{-}\right\rangle $
govern the carrier excitation.

In summary, based on the entropy production rates of far-from-equilibrium
open systems, we prove that the energy conversion performances of
these systems can be improved by suitably adjusting the coupling strengths
between the energy and matter currents. We apply the theroy to a DQD
sytem that is weakly coupled to two separate electronic reservoirs
and attain its performance upper bound. The results show that the
universal thermodynamic coupling rule can open a new feld in the application
of nano-devices.

\begin{acknowledgments}
We thank Professor Gernot Schaller for helpful discussions. This work
has been supported by the National Natural Science Foundation of China
(Grant No. 11805159) and the Fundamental Research Fund for the Central
Universities (No. 20720180011).
\end{acknowledgments}

\end{document}